\documentclass[fleqn,10pt]{wlscirep}
\usepackage[utf8]{inputenc}
\usepackage[T1]{fontenc}

\makeatletter
\@fpsep\textheight
\makeatother

\title{Quantifying yeast colony morphologies with feature engineering from time-lapse photography}
\author[1,*]{Andy Goldschmidt}
\author[2,*]{James Kunert-Graf}
\author[2]{Adrian C. Scott}
\author[2, 3]{Zhihao Tan}
\author[2, 3]{Aim\'{e}e M. Dudley}
\author[4]{J. Nathan Kutz}

\affil[1]{Department of Physics, University of Washington, Seattle, WA 98195, USA}
\affil[2]{Pacific Northwest Research Institute, Seattle, WA, 98122, USA}
\affil[3]{Molecular and Cellular Biology Graduate Program, University of Washington, Seattle, WA 98195, USA}
\affil[4]{Department of Applied Mathematics, University of Washington, Seattle, WA 98195, USA}
\affil[*]{corresponding authors: Andy Goldschmidt (andygold@uw.edu), James Kunert-Graf (jkunert@pnri.org)}
\begin{document}

\begin{abstract}
Baker's yeast (\textit{Saccharomyces cerevisiae}) is a model organism for studying the morphology that emerges at the scale of multi-cell colonies. 
To look at how morphology develops, we collect a dataset of time-lapse photographs of the growth of different strains of \textit{S.~cerevisiae}.
We discuss the general statistical challenges that arise when using time-lapse photographs to extract time-dependent features. 
In particular, we show how texture-based feature engineering and representative clustering can be successfully applied to categorize the development of yeast colony morphology using our dataset.
The \textit{Local binary patterns} (LBP) from image processing is used to score the surface texture of colonies. 
This texture score develops along a smooth trajectory during growth. The path taken depends on how the morphology emerges.
A hierarchical clustering of the colonies is performed according to their texture development trajectories. 
The clustering method is designed for practical interpretability; it obtains the best representative colony image for any hierarchical sub-cluster.
\end{abstract}

\flushbottom
\maketitle

\thispagestyle{empty}

\section*{Background \& Summary} \label{sec:background}
\textit{Saccharomyces cerevisiae}, or Baker's yeast, is a model organism that can grow a diverse range of strain-dependent morphologies at the scale of cell colonies~\cite{kuthan2003domestication,vopalenska2010role,vachova2011flo11p}. Manifesting as macroscopic tubes, cracks, and ridges that mark the colony surface, these morphologies influence how the organism interacts within a biological environment~\cite{costerton1999bacterial,parsek2005sociomicrobiology}. Beer brewers and bakers have historically relied on visual identification of morphology to make practical distinctions among encountered strains of \textit{S.~cerevisiae}~\cite{hall1971detection, spencer2013yeasts}. 

Easy access to digital cameras and video recording allows for the eyes of a well-trained expert to be replaced with computer algorithms well-trained on large collections of image data. The data can be brought to bear on scientific questions of which yeast morphology identification is only one emblematic example. The challenge of this broad paradigm shift toward data-driven science is how to leverage statistical methods and algorithms to translate the abundant \textit{big data} into useful information~\cite{brunton2019data}. This is especially pertinent when navigating the high-dimensional pixel spaces associated with digital photographs and video.

The dataset introduced by this article involves large ensembles of \textit{S.~cerevisiae} growth experiments photographed hourly over days.  We provide a suite of statistical methods to translate these time-lapse images into actionable information for understanding colony morphology. Although our analysis is illustratively applied to the specific question of yeast colony morphology, our approach can be understood more broadly as a strategy for quantitatively extracting desired features from time series of photographs. Toward this end, we emphasize in our Methods section the general challenges faced in the analysis of time-lapse image data. We review alternative approaches using the literature on colony morphology as a guide, and explain the general benefits of the perspective introduced by this paper. All of our statistical methods are implemented with generic Python packages prioritizing easy adaptation to any time-lapse image datasets.

\begin{figure}[!htbp]
\centering
    \includegraphics[width=0.8\textwidth]{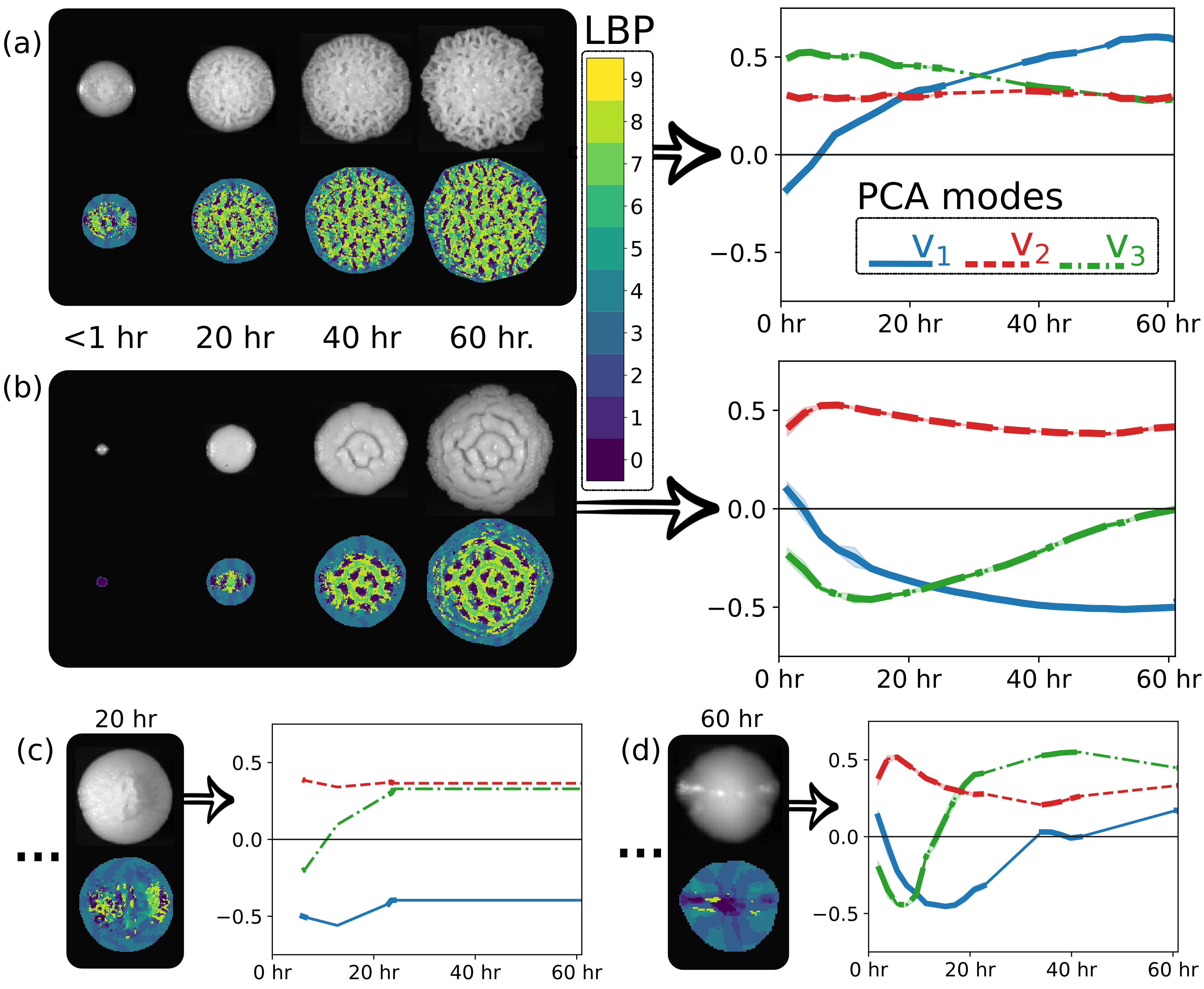}
    \caption{Figures~\ref{fig:main}(a)-(d) show the result of feature engineering for some colonies in the dataset. Later in Figure~\ref{fig:cluster} these colonies are shown to be the top level prototypes returned by the hierarchical clustering algorithm. We will illustrate how to read this figure by taking Figure~\ref{fig:main}(a) as an example. In the left block, the colony image and the local binary pattern (LBP) categorization of the colony image are compared at beginning, middle, and end times measured in hours.  In the right block, the images have been projected onto the three Principal Component Analysis (PCA) modes used for dimensionality reduction of the images at all measurement times. The corresponding feature space trajectories are plotted.  Bold lines correspond to existing data that has been binned. Thin lines are missing values filled in using interpolation of the existing data. Distances between the trajectories shown here are used to define the pairwise distance matrix used for clustering.}
    \label{fig:main}
\end{figure}

Even in scientific settings, past categorization methods of yeast colony morphology have relied on qualitative scoring in which a single investigator identified colony categories by eye~\cite{granek2010environmental,vst2010general,voordeckers2012identification}. Additional tools such as ImageJ~\cite{schneider2012nih} and CellProfiler~\cite{lamprecht2007cellprofiler} have been used to identify colonies based on shape, size, and color. Another approach was categorization based on principal component analysis at fixed time points~\cite{ruusuvuori2014quantitative}. For time-series of images, size trajectories have been studied for characterizing morphology~\cite{memarian2007colony, dittmar2010screenmill}.  In this article, we demonstrate the unsupervised categorization of morphology based on complete texture trajectories as shown in Figure~\ref{fig:main}. Our contribution can be understood in two ways. First, by introducing the texture analysis tools from image processing we demonstrate feature engineering appropriate for the practical quantification of morphology. Second, our approach enables the study of time series of images because of the smooth trajectories of the images once projected into our engineered feature space. From this perspective, this article is a tutorial on a general framework enabling the principled categorization of datasets involving time-lapse image data.

In particular, we introduce a novel statistical framework combining texture analysis tools from image processing with clustering algorithms from machine learning. All of the necessary tools for the analysis pipeline of the framework are readily available in Python (refer to Code Availability). In addition, a supplementary open-source Python package has been introduced by the authors for the purpose of improving upon the available clustering algorithms~\cite{goldschmidt2021pyp}. The supplemental provides a Python implementation of a hierarchical clustering algorithm that finds a prototypical representative for each data cluster it obtains~\cite{bien2011hierarchical}. Prototypical representatives of clusters are practical aides especially relevant for large image-based datasets. This article will use its photographic dataset of \textit{S.~cerevisiae} colonies to demonstrate that this novel framework combining texture analysis and machine learning provides a simple, interpretable, and effective approach for the categorization of colony morphologies.

\section*{Methods} \label{sec:methods}
This section describes (i) the experimental design for image data acquisition, (ii) image pre-processing, (iii) texture-based feature engineering, and (iv) representative clustering.  These steps are part of the characterization pipeline as shown in Figure~\ref{fig:main}. All code used in these methods are available in standard open-source Python libraries or are provided by the authors as fully-documented Python packages downloadable from the Python Package Index (see Code Availability).

\subsection*{Yeast strains and genetic manipulation} \label{sec:strains}
Unless noted, standard media and methods were used for the growth and manipulation of yeast~\cite{sherman1987methods} The \textit{S.~cerevisiae} strains used in this study are listed in Table~\ref{tab:strains}. Strains with a YPG prefix, which are included in the image dataset, are the haploid progeny of two yeast strains.  The first strain, YO502, was derived from an Ethiopia white tecc brewing strain~\cite{sirr2018natural}. The second strain, YO1817, is isogenic to the YO486 Japanese sake brewing strain~\cite{sirr2018natural}, except that it is monosomic for chromosome I and disomic for chromosome XII. YO502 and YO1817 were mated, the heterozygous diploid was sporulated, and haploid recombinant progeny were isolated by manual tetrad dissection.

\subsection*{Image data acquisition} \label{sec:experiment}
Colonies of \textit{S.~cerevisiae} were generated by arraying approximately 48 individual cells of the same strain onto an agar plate in a checkerboard pattern using a Sony SH800 fluorescent cell sorter with 488~nm and 531~nm excitation lasers. Photographs were acquired while these colonies grew at $30^\circ$C in rich (YPD) medium ($1\%$~yeast extract, $2\%$~peptone, $1.5\%$~glucose). Colony images were acquired at median rate of one image every $23$~minutes over the course of $3$~days using a Canon 5d Mark II camera outfitted with a MP-E 65 mm 1-5x macro lens. This camera was attached to a custom built motorized 2-axis gantry that moved the camera over a set of up to 13 plates. For this particular study, 5500 colonies were selected as a representative ensemble of experimental data. These colonies represent 196 unique strains of the species. More information about this dataset is discussed in the Data Records section. Access to a larger library of data from which this dataset is sampled can also be made available upon request. Contact the corresponding author(s) for more information.

\subsection*{Image pre-processing} \label{sec:preprocessing}
To be studied further, colony images must be extracted from the image backgrounds. Backgrounds included agar plate edges or segments of other colonies in the grid. Image selection was accomplished with an image processing pipeline using algorithms in the scikit-image library for Python~\cite{vanderWalt2014}.

The pre-processing  pipeline involved three main steps. (i) Canny edge detection (\textit{skimage.feature.canny}) determined the edges and boundaries of any objects in the image. This included the colony boundary. (ii) The circular Hough transform (\textit{skimage.transform.hough\_circle\_peaks}) found the maximal radius of the largest (approximately) circular object in the Canny edge detection image. This was assumed to encompass the colony boundary. Anything outside this circle was masked. (iii.) A convex hull (\textit{skimage.morphology.convex\_hull\_image}) was drawn around the unmasked region of the original image to capture the true (non-circular) colony boundary in detail. Parameters in these functions were tuned by hand to the data. Finally, the high-resolution images were re-sized to one-half of the approximate average pixel dimensions, $100 \times 100$. Images were padded before re-sizing to preserve colony shapes.

\subsection*{Feature engineering using local binary patterns} \label{sec:feature}
The dominant visible morphology of a growing yeast colony is the unique and complicated pattern of folds and lumps that evolve primarily on the colony surface. This texture is an ensemble of 2-dimensional coherent spatiotemporal structures. Dimensional reduction of this kind of complex texture space is important for discovering the appropriate descriptors to distinguish the available morphologies. 

A de facto standard for dimensionality reduction is the Principal Component Analysis (PCA)~\cite{jolliffe2016principal}. Given an image dataset, PCA finds the minimal set of basis images that inform the majority of the variability in the dataset. If a PCA algorithm was applied to the yeast image dataset, one might expect a set of 2-dimensional static images describing the full variety of colony image textures. Unfortunately, a well-known issue with PCA is that even the simplest dynamic structure requires a large set of static basis images to accurately describe its motion. For example, a movie of a bouncing ball travelling across a fixed field of view would need an entire flip-book of static images to show the traversal. When applied to growing yeast colonies, PCA is being asked to use static basis images to simultaneously describe the patterns of folds and lumps on a colony surface as well as the outward-bound radial traversal of those patterns. The generic growth information is dominant and will win out at the expense of the more subtle distinctions between texture patterns. This discussion is intended to communicate why the naive application of standard dimensional reduction via the PCA in the image space is not \textit{robust to growth}; distinctions among morphologies are lost.

Two straightforward options remain. Either (1) remove time altogether, or (2) engineer better features for texture that are robust to growth. As to the former option, Reference~\cite{ruusuvuori2014quantitative} showed how yeast colonies could be clustered according to their two-dimensional image at the final recorded time. A disadvantage of using a fixed time is this ignores a lot of the available information in our dataset. Hence, the analysis in this article follows the latter option. We engineer accessible and appropriate features for yeast texture via a standard image processing tool called the local binary pattern (LBP) \cite{ojala2002multiresolution}. We will show that the LBP is robust to growth. A Python implementation of the LBP algorithm is available in scikit-image as \textit{skimage.feature.local\_binary\_pattern} allowing for ready accommodation to our data processing pipeline.

\begin{figure}[!htbp]
    \centering
    \includegraphics[width=\textwidth]{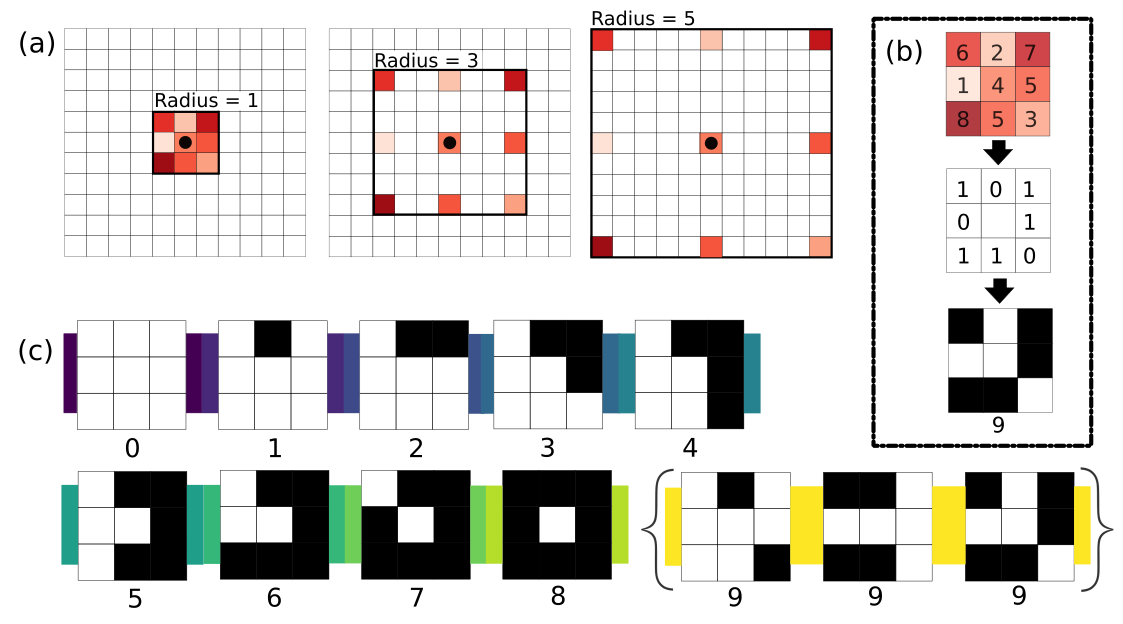}
    \caption{Figure~\ref{fig:lbp}(a) shows the radius parameter of the local binary pattern (LBP) is adjusted to select the neighboring pixels for the algorithm. In Figure~\ref{fig:lbp}(b), the LBP of one pixel is calculated. Neighboring pixels are judged according to the center pixel and the result is categorized. Figure~\ref{fig:lbp}(c) shows the nine LBP categories. In Category 0, all eight neighbors are bright. In Category 1, all neighbors are bright except for a single pixel that is dark, and so on. Category 8 is a dark spot.  Any discontiguous blend of light and dark pixels falls within the catch-all Feature 9.}
    \label{fig:lbp}
\end{figure}

The LBP was designed to capture the essential notions of texture in 2-dimensional images. For example, texture should not depend on the lighting or orientation at which the original image was viewed--therefore, the LBP is scale and rotationally invariant by its design. The LBP algorithm uses ten pattern categories to label each pixel in an image. The pattern category for a pixel is chosen by comparing that pixel to eight of its surrounding neighbors (Figure~\ref{fig:lbp}). We can apply the LBP to every pixel in an image from our colony image dataset. Next, we can record the relative frequency of each LBP-pattern category amongst all of the image's pixels. This produces a vector of 10 numbers which we can associate to that image. In this article, we refer to this engineered feature space of vectors describing our colony images as the LBP space. Importantly, the LBP space has no spatial information. The spatial information was integrated out by computing the relative frequencies of the LBP-pattern categories for each image. This is the critical step for making our LBP feature robust to growth. We have replaced the complicated image of texture--many folds and lump evolving together on a 2-dimensional surface--with a simple trajectory that changes smoothly in the 10 coordinates that define the LBP-space.

Because LBP-space trajectories are robust to growth, dimensionality reduction can be applied in this feature space. A PCA algorithm was used to determine the best set of static basis modes for describing the full variety of our dataset in this 10-dimensional LBP-space. Each PCA mode returned by the algorithm was a vector of 10 numbers. Taken together, these LBP PCA modes provided an optimal set of coordinates to describe the LBP-space trajectories. We found that 3 LBP PCA modes were sufficient to characterize a majority of the variation of the LBP-space trajectories. Hence, we ultimately described our photographed morphology growth using this reduced 3-dimensional feature space.

We now make a few practical comments regarding the robustness advantages of this feature space. First, in the provided dataset our photographs of yeast colonies were taken at inconsistent times. The lack of uniform measurement time is due to the sporadic traversal of the camera during the experiment. Favorably, the smooth trajectories in the LBP space allowed for us to use interpolation so every colony shared the same measurement times. This interpolation could not have been done in the full image space. In our study, we also simplify by setting a universal final time which was fixed to be the median ending time of our dataset. The median is robust to short-time outliers involving failed-growth experiments. If an experiment ended earlier than our chosen final time, growth was assumed to have saturated and the final data point was extrapolated.

\subsection*{Representative clustering} \label{sec:clustering}
We categorized colony texture within the 3-dimensional feature space defined by the LBP PCA modes. To do so, model-free clustering was applied to the feature-space trajectories of the colonies. Model-free emphasizes that distances were defined directly between trajectories without fits or other modelling assumptions. To define colony distances, the usual Euclidean distance was used to compute a distance between points in the 3-dimensional feature space at each time. Summing this distance over all times in the trajectory gave us a valid distance metric to use for the yeast colonies.

There are a wide variety of machine learning methods for finding clusters once pairwise distances have been obtained; in this paper, hierarchical clustering \cite{hastie2009unsupervised} was used because of its natural emphasis on differences at various scales. Distinct clusters at the top of the hierarchy can be expected to have obvious morphological differences. Meanwhile, differences amongst clusters making up the lower rungs of the hierarchy are expected to have more subtle contrasts. The standard implementation of agglomerative hierarchical clustering assembles clusters from the bottom up according to some linkage rule for joining pairs of clusters. A variety of Python algorithms for hierarchical clustering are found in e.g. SciPy \cite{virtanen2020scipy}.

The output of a hierarchical clustering algorithm is a binary tree called a dendrogram. In a dendrogram, the leaves (made up of the initial data points) are nodes placed at height zero. Linked clusters are drawn as new nodes above their constituent pair and are placed at a height proportional to the linkage distance. The linkage distance is the quantity the hierarchical clustering algorithm uses to decide which cluster to join, and the definition can vary between implementations~\cite{hastie2009unsupervised}. The assembly proceeds until the final linkage when the binary tree is crowned by one last node. 

The dendrogram can be cut at different heights. A cut is a choice of partition of the initial data points according to the sub-trees appearing directly below the cut. For our dataset, we would like for the dendrogram cut height to say something interpretable about the differences between morphology among the partitions. That is, higher up in the tree we expect the differences to be more dramatic.  Lower in the tree we expect the differences to be more subtle. In an image-based dataset, we have the opportunity to introduce additional interpretability. For each cluster, we can select an optimal representative image. The representative image should offer a best approximation to the shared yeast morphology approximated by the cluster.  Of the many linkage rules in standard practice~\cite{hastie2009unsupervised,virtanen2020scipy} (summarized in Table~\ref{tab:linkages}), minimax linkage\cite{bien2011hierarchical} was the only choice that met the desired criteria of interpretable cuts and natural cluster representatives important for our image-based dataset.

\begin{figure}[!htbp]
    \centering
    \includegraphics[width=\textwidth]{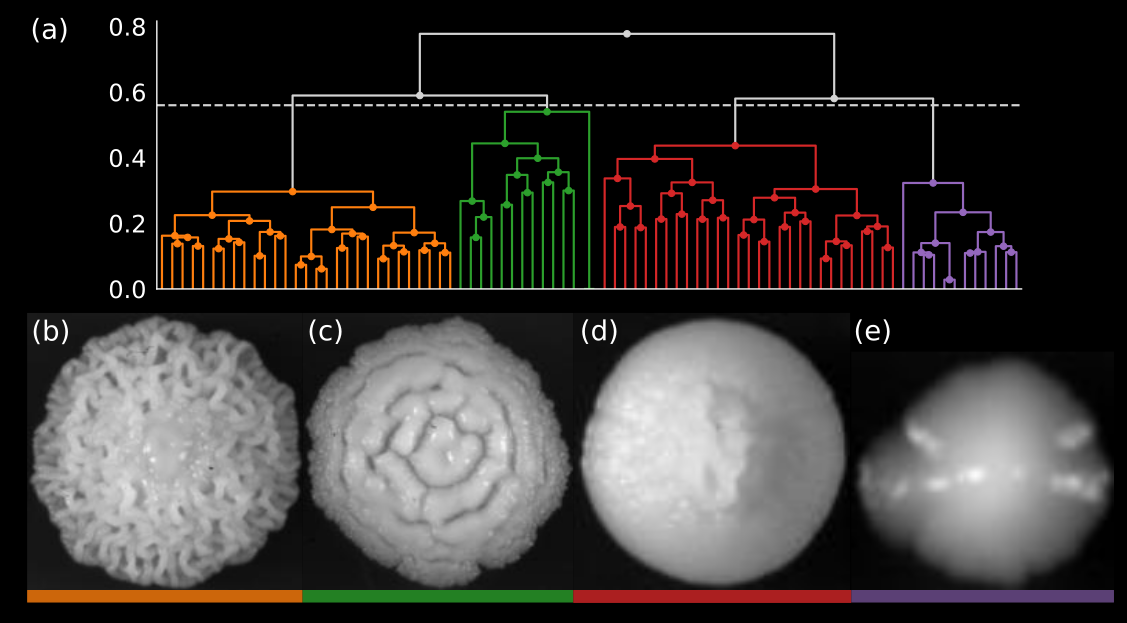}
    \caption{Figure~\ref{fig:cluster}(a) shows the dendrogram constructed by applying agglomerative hierarchical clustering using minimax linkage from the pairwise distance matrix defined by the feature engineering depicted in Figure~\ref{fig:main}. Four dendrogram branches have been distinguished by a fixed-height cut of the dendrogram (dotted line). From left to right, these four branches correspond to the cluster prototypes shown in Figures~\ref{fig:cluster}(b)-(e). Clustering is based upon the entire time-lapse trajectory. The displayed images are of each colony at the latest available time in its time-lapse: 58 hr., 60 hr., 24 hr., and 58 hr. for Figures~\ref{fig:cluster}(b), (c), (d), and (e).}
    \label{fig:cluster}
\end{figure}

The result of hierarchical clustering using minimax linkage is shown in Figure~\ref{fig:cluster}. The upper section, Figure~\ref{fig:cluster}(a), shows a dendrogram with a view that has been truncated to show only the last seven linkages computed by the algorithm. We do this to avoid viewing the full dendrogram which must eventually display a node for each point in the dataset. In Figures~\ref{fig:cluster}(b)-(e), four prototype images are shown and the growth time is displayed below the image. The four prototypes which display obvious differences in texture are the optimal representatives of the four top-level clusters produced by the cut dendrogram. The cut occurs at a height that has been chosen according to the CH index. The CH index~\cite{calinski1974dendrite} is a standard statistical measure balancing scores for variations within and between clusters. The index is a useful metric for selecting appropriate cut heights.

\begin{figure}[!htbp]
    \centering
    \includegraphics[width=\textwidth]{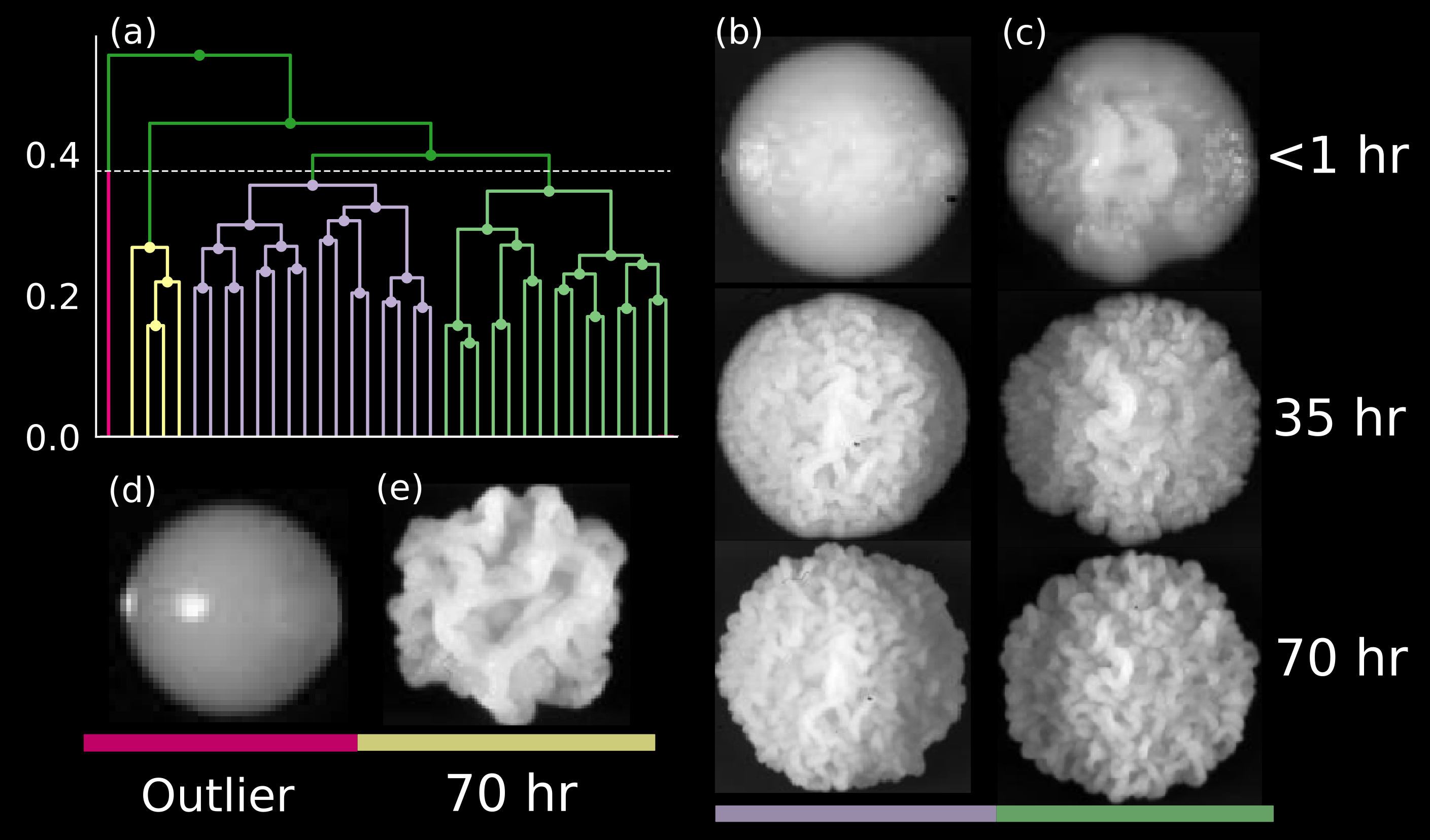}
    \caption{Figure~\ref{fig:subcluster}(a) shows the sub-branch represented by the prototype Figure~\ref{fig:cluster}(c) of the dendrogram plotted in Figure~\ref{fig:cluster}(a). A new cut height is chosen (represented by a dotted line in Figure~\ref{fig:subcluster}(a)) and four branches are created. The four prototypes in Figures~\ref{fig:subcluster}(b)-(e) ordered alphabetically correspond to the four branches read left to right. Numbers next to the colony images report the elapsed time of the experiment (in hours) when that image was taken. Figure~\ref{fig:subcluster}(d) shows an outlier colony for which no growth data was recorded during the experiment (less than 1 hour had elapsed). Figures~\ref{fig:subcluster}(b),(c) show colony growth at three times along the time-lapse to demonstrate how the complete texture trajectories are used to define the colony distances.}
    \label{fig:subcluster}
\end{figure}

The advantage of hierarchical clustering is demonstrated by Figure~\ref{fig:subcluster}. In Figure~\ref{fig:subcluster}(a), the sub-dendrogram extending below the branch represented by Figure~\ref{fig:cluster}(c) is shown with a new (lower) cut height. As such, Figures~\ref{fig:subcluster}(b)-(e) show prototypes with more subtle contrasts than those evident among the previous Figures~\ref{fig:cluster}(b)-(e). Exploring the dendrogram tree obtained from minimax linkage allows for visual insights into the structure of texture-based clusters. Agglomerative hierarchical clustering is able to provide texture information at multiple scales. Additionally, Figures~\ref{fig:subcluster}(b),(c) show how subtle distinctions can manifest in texture growth at early times--this additional contrast cannot be obtained by comparing only the terminal colony textures.

\subsubsection*{Code Availability} \label{sec:code}
All algorithms used in this paper can be freely obtained from open source Python packages.

The image pre-processing pipeline and feature engineering (Canny edge detection, circular Hough transform, convex hull, local binary pattern) were performed by using the algorithms implemented in the scikit-image library~\cite{vanderWalt2014}. The Principal Component Analysis was carried out with the Singular Value Decomposition using the numerical python library, NumPy~\cite{harris2020array} (\textit{numpy.linalg.svd}). Hierarchical clustering tools made available by SciPy\cite{virtanen2020scipy} in the module \textit{scipy.cluster.hierarchy} were also used.

The algorithm for minimax linkage has been published as an R package called \textit{protoclust}~\cite{Bien2019}. Independently, the authors of this paper developed the \textit{pyprotoclust} Python package~\cite{goldschmidt2021pyp} (albeit with the retroactive incorporation of this previously established naming convention). Our Python implementation allows us to obtain seamless integration with the interface established by SciPy's hierarchical clustering and the other Python-specific tools used in this study. Our implementation can also take advantage of multi-threading to improve algorithm performance. The open-source code for the \textit{pyprotoclust} Python package can be obtained from the the Python Package Index under the MIT license. The documentation is hosted publicly online by Read~the~Docs (\href{https://pyprotoclust.readthedocs.io/en/latest/}{pyprotoclust.readthedocs.io}).
 .

\section*{Data Records} \label{data}
There are four data records accompanying this report. These four data records (\textit{ImageData}, \textit{CoordinatesTable}, \textit{DistancesTable}, and \textit{StrainsTable}) are hosted on figshare~\cite{goldschmidt2021fig}.

The first data record \textit{ImageData} is a 3.4 GB zip file of yeast colony images stored as JPEG files hosted online~\cite{goldschmidt2021fig}. The files are contained in folders named for the inception date of that experiment and the camera used to produce the images e.g. 2014-10-23-Cam1. There are 18 folders in total. Inside each folder are tens of thousands of image files with names like YPG11407\_001\_1080.jpg; we can break these names down into easy to understand pieces. First, the YPG refers to the growth medium of the yeast colony: in this case, for Yeast Extract–Peptone–Glycerol (YPG) Agar. Second, the number 11407 refers to the strain (consult data record \textit{StrainsTable}). Third, the number 001 identifies the strain experiment iteration because many redundant strains were grown. Finally, 1080 is the photograph timestamp measured in minutes.

The second data record \textit{CoordinatesTable} is a {CSV} table hosted online~\cite{goldschmidt2021fig} for use in e.g. the Python database library Pandas~\cite{mckinney2010data, reback2020pandas}. An example row is given in Table~\ref{tab:pca}. The full table contains the LBP PCA coordinates for each image (refer to the Methods section, under Feature engineering using local binary patterns). The relative path and image file from data record \textit{ImageData} are included as the first two columns to allow for identification with images. For simplicity, the timestamp of the photograph is separated from the filename and included as a separate column. Finally, the values of the yeast colony image as measured in the coordinates of the 3 LBP PCA modes are provided. This table will allow the user to by-pass the data processing pipeline for image extraction and LBP calculation needed to produce the LBP PCA modes used for dimensionality reduction and trajectory analysis.

The third data record \textit{DistancesTable} is a {CSV} table hosted online~\cite{goldschmidt2021fig}. An example row is given in Table~\ref{tab:distance}. The full table contains the distance matrix discussed in the Methods section under the heading Representative clustering . The distance matrix lists the pairwise distances between unique iterations of colony experiments; identification is by a unique number and a corresponding Root Filename like YPG11407\_001 which stands in for the collected image files matching YPG11407\_001\_*.jpg where the time values have been replaced by the wildcard symbol, *.  The data matrix is presented in a vector format. In the vector format, we do not store the entire distance matrix. We only include the necessary list of index pairs $(i,j)$ because a distance matrix is symmetric, $\textrm{distance}(i,j) = \textrm{distance}(j,i)$. The first two columns are the pair of indices between which the distance is defined. The third and forth column are the Root Filenames associated with these indices. The fifth column is the distance value. The distance values are the mean (integrated) distance between interpolated LBP PCA trajectories (see the Methods section under Feature engineering using local binary patterns). Converting between the provided vector format for the distance matrix and the square distance matrix format is straightforward: in Python, consult the documentation for the \mbox{scipy.spatial.distance.squareform} method.

The fourth data record \textit{StrainsTable} is included in the paper as Table~\ref{tab:strains}. It contains the genotype and source for each yeast strain appearing in our data.

\section*{Technical Validation}
In this section, we emphasize two points about the technical quality of the dataset of yeast colony images. First, we explain the reliability of the colony metadata. The metadata refer to the folder and file names in the data record called \textit{ImageData}. We also make some comments on outlier detection. Second, we discuss why it is reasonable to conclude that growth conditions are approximately the same for all the colonies. Equal growth conditions are an important assumption when we create our basis of feature-space growth trajectories (see the Methods section). 

The essential metadata for each image file in the \textit{ImageData} data record are the strain number, the strain experiment iteration, and the photograph timestamp. For additional information on the manifestation of this information within the file, consult the Data Records section. The most straightforward metadata entry was the timestamp which we extracted directly from the metadata of the digital photograph itself. Next, consider the strain number. All of the single colonies used in this study are replicates. Each agar plate was entirely filled by colonies of the same strain. With this redundancy, we can conclude that the strain number was reliable even if a single strain experiment iteration exhibited an uncharacteristic morphology. To address the reliability of the strain experiment iterations, we must recall some facts of the experiment design. Strain experiment iterations were arranged in a checkerboard pattern on a plate. A digital camera was attached to a custom built motorized 2-axis gantry that allowed the camera to move over the iterations. A computer program managed and tracked the motorized control of the camera. The tracking location and time could be compared with the image timestamp to correctly associate an image with its strain experiment iteration.

Even with these reliable designs in place, let us suppose a colony were mislabelled at some point during its growth. We can expect that the new image with the incorrect label will have poor correlation with the previously photographed image which had correctly been assigned this same label. Indeed, time correlations measurements like this are are often used, e.g. to identify noise in gene expression profiles~\cite{kruglyak2001new}. In our Methods section, we discuss clustering the photograph data using a texture score that develops along a smooth trajectory during growth. An important benefit of the proposed approach is that the smoothness of the trajectory emerges from the continuity of the colony growth. Mislabelling breaks the continuity. Hence, we naturally detect mislabellings as outliers in the hierarchical clustering results. As a final comment, we note that if the label is incorrectly assigned early in the growth like at the first timestep, there is not much that can be done to detect abnormalities. Care must be taken that the labels have been assigned correctly for the initial pass of the camera.

Next, we address the reliability of assuming that colonies develop under equal growth conditions. Again, recall that the colonies are arranged in a checkerboard pattern. This pattern was designed with appropriate spacings to grant equal nutrient access regardless of the colony position. All colonies were grown under the same conditions: at $30^\circ$C in rich (YPD) medium ($1\%$ yeast extract, $2\%$ peptone, $1.5\%$ glucose). The special-purpose camera rig was built within the growth environment. The camera rig was located at a distance from the colonies and run at a sufficiently slow speed to mitigate any undesired heating.

\section*{Usage Notes}
A tutorial has been written for reproducing the pre- and post-processing results discussed in this report. This tutorial is implemented as an interactive Jupyter notebook~\cite{kluyver2016jupyter} and can be obtained on GitHub at~\url{https://github.com/pyYeast/yeast-morphology-tutorial}. Self-contained instructions are included with the tutorial. Questions about the tutorial can be directed to the repository manager(s) through GitHub.

\bibliography{main}


\section*{Author contributions statement}
A.D. conceived the experiment. A.S., Z.T., and A.D. conducted the experiment. A.G., J.K., and N.K. analyzed the results.  A.G. drafted the manuscript, and all authors reviewed it.

\section*{Competing interests} 
A.D. is a scientific advisory with a financial interest in Fenologica Biosciences, Inc. (mandatory statement)

\newpage
\section*{Figures \& Tables}

\begin{table}
  \caption{\textit{S.~cerevisiae} strains used in this study.}
  \label{tab:strains}
  \centering
  \begin{tabular}{lll}
    \toprule
    Strain name & Genotype & Source \\
    \midrule
    YO502   & MAT$\alpha$ ho$\Delta$0::hphMX6, SPS2:EGFP:natMX4    & Reference~\cite{sirr2018natural} \\
    YO486   & MAT\textbf{a} ho$\Delta$0::hphMX6, SPS2:EGFP:kanMX4, ser1-S, Disomy I     & Reference~\cite{sirr2018natural} \\
    YO1817  & MAT\textbf{a} ho$\Delta$0::hphMX6, SPS2:EGFP:kanMX4, ser1-S, Disomy XII   & This study \\
    YPG11406-YPG11693 & Haploid progeny of the YO502 $\times$ YO1817 cross & This study \\
    \bottomrule
  \end{tabular}
\end{table}
 
\begin{table}
  \caption{Comparison of common linkage functions}
  \label{tab:linkages}
  \centering
  \begin{tabular}{lll}
    \toprule
    Linkage     & Interpretable cut & Natural cluster representatives\\
    \midrule
    Single      & Yes               &  No   \\ 
    Complete    & Yes               &  No   \\ 
    Average     & No                &  No   \\
    Centroid    & No                &  Yes  \\ 
    Minimax     & Yes               &  Yes  \\
    \bottomrule
  \end{tabular}
\end{table}

\begin{table}
    \caption{An example line of the data record \textit{CoordinatesTable}.}
    \label{tab:pca}
    \centering
    \begin{tabular}{llllll}
        \toprule
        Folder &  Filename &   Time    &   PCA mode 1  &   PCA mode 2  &   PCA mode 3 \\
        \midrule
        {2014-10-23-Cam1} & {YPG11407\_001\_1080.jpg} & $1080$ & $-0.077481$ & $0.472005$ & $0.372762$ \\ 
        \bottomrule
    \end{tabular}		
\end{table}

\begin{table}
    \caption{An example line of the data record \textit{DistancesTable}.}
    \label{tab:distance}
    \centering
    \begin{tabular}{lllll}
        \toprule
       I & J & I Root Filename &  J Root Filename &  Distance \\
        \midrule
        0 & 3132 & {YPG11407\_001} & {YPG11565\_20} & 0.347 \\ 
        \bottomrule
    \end{tabular}		
\end{table}

\end{document}